# CardSharp: A python library for generating MCNP6 input decks

Nikhil Deshmukh,[a]* and Mital Zalavadia[a]

[a]*Pacific Northwest National Laboratory, Richland, Washington*

*E-mail: nikhil.deshmukh@pnnl.gov

# CardSharp: A python library for generating MCNP 6 input decks

A python library for the creation of MCNP6 input decks is described. The library supports geometry generation with automatic assignment of surface/facet numbers, cell numbers, transform numbers and material numbers along with MCNP Universes and FILL feature. Rectangular and Hexagonal Lattices are also supported. A large material library is included. Support for a good selection of common sources and tallies is also provided. Cards or features which are currently not supported in the library can also be inserted as raw strings into the output stream. Combining Python features like descriptively named variables, functions and for loops with library functions provides an intuitive and parametric way to create, modify and maintain complicated geometries and simulation models. The generated card deck also has human readable comments which makes it easy to read and relate back to the python source. Some support for running MCNP, reading tallies and plotting is also provided.

**I. INTRODUCTION**

MCNP[1] is a very popular and powerful tool used for modeling radiation transport problems. Over the years, it has acquired the status of being the gold standard in radiation modeling and has very powerful variance reduction techniques. However, the MCNP inputs decks can be complicated, hard to organize, and therefore time consuming to generate. The format was probably considered manageable when MCNP was first released but would not be considered user friendly today. Besides being time consuming, it is also prone to syntax errors. The motivation behind developing the Python library described here is to modernize and speed up input deck creation, reduce formatting errors, and take out the guess work that a novice MCNP user might face. Experienced users will also find it useful for generating complicated decks because using the Python language allows for descriptive names and parameterized generation of part dimensions and transformations.

**II. SUPPORTED FEATURES**

*1. Geometry*

MCNP geometry definition consists of two steps: defining surfaces and defining cells. There are two slightly different ways to define surfaces: Using primitive surfaces and using macro bodies. The library described here has good support for both. Part of the geometry definition is to assign a material to each of the cells and the library supports this, too. Global importance card (IMP) and per-cell importance strings are also supported. The generated file can be visualized with a tool such as Gxsview[2]or Vised[3] to create a streamlined process for generating geometries.

The following geometry cards are supported: All ten macros are supported (ARB, BOX, RPP, SPH, RCC, TRC, ELL, RHP/HEX, TRC, WED). All surfaces like plane (P, PX, PY, PZ), cylinder (C/X, C/Y, C/Z), cone (K/X, K/Y, K/Z), torus (TX/TY/TZ), Special Quadratic (SQ), General quadratic (GQ), surface-defined-by-points (XYZP) are supported (other than some redundant options). Multiple universes, (UNI, FILL option) and rectangular/hexagonal lattices (LAT) are also supported.

The calls to insert surfaces/macros return the auto assigned surface/macro numbers which can be used symbolically in later sections to insert cells comprising of unions/intersections of the surfaces. Cell numbers can be chosen manually or auto assigned. The auto assigned cell numbers can be used later to instantiate sources and tallies with the appropriate cell numbers. Support is also provided for defining the world and graveyard, either using automatic cell complement which works well for smaller problems or with a user supplied surface list which can be computationally more efficient.

*2. Materials*

More than 400 materials are already included in the library in the form of a Python dictionary. (PNNL materials compendium [4]) The user can include the needed materials into their input deck by using the appropriate key for the material or creating an easier to use alias. The dictionary contains a default density for each material which can be overridden if needed. More materials including dummy materials needed for certain problems can be introduced directly from the python code using the MCNP material syntax, with automatic assignment of material numbers.

*3. Sources*

The following types of sources are supported.

1. Point source, isotropic – User provides source position, energy distribution, distribution type, and particle type.

2. Point source, mono directional – User provides source position, direction vector, energy distribution, distribution type, and particle type.

3. Point source, with angular biasing/restriction – User provides source position, direction vector, cone angle, energy distribution, distribution type, and particle type.

4. Spherical volume source, with angular biasing/restriction - User provides position, radius, direction vector, cone angle, energy distribution, distribution type, particle type, and rejection cell.

5. Cylindrical volume source, with angular biasing/restriction – User provides position, radius, axis, thickness, direction vector, cone angle, energy distribution, distribution type, particle type, and rejection cell.

6. Box volume source, with angular biasing/restriction – User provides x/y/z range of the box, direction vector, cone angle, energy distribution, distribution type, particle type, and rejection cell.

Sources that are too complicated or too specific for a particular application can be inserted as raw text from within the python script so that the benefits of using the CardSharp library for other sections can still be realized.

4. *Tallies*

Support for the following tally types is included.

1. F1, F2, F4, F6, F7, F8 – User provides tally number, surface/cell info, particle type, energy list, and multiplier list.

2. F5 – User provides tally, position, exclusion radius, particle type, energy list, and multiplier list.

3. FIR5 – User provides position, normal to the radiography tally, number of bins in each of the two dimensions, energy list, multiplier list, and particle type.

Support for special treatment cards is in the works but can currently be introduced into the deck using the general function insertIntoDataSection which takes a raw multi line MCNP string as input.

Tallies that are too complicated or too specific to a particular application can also be inserted as raw text from within the Python script so that the benefits of using the CardSharp library for other sections can still be realized.

5. *Physics and Output control cards*

Physics card support for NOCOH, IDES, NODOP, CUTN, CUTP, CUTE is included. Support for number of particles (NPS), PRDMP, NOTRN cards is also included.

6. *Running, parsing results*

Support for running MCNP on the local machine with the generated deck is provided. A default MCNP installation path is included, which can be changed if needed. Limited supported for reading mctal tally files is also provided so that the full model development cycle can be done from within Python. Future support for execution on a High-Performance Computing (HPC) cluster is also planned.

**III. ILLUSTRATING USAGE OF LIBRARY FUNCTIONS USNIG CODE SNIPPETS**

Here we will give a flavour of how the library works by going over some of the most commonly used library methods. There are more elaborate examples on the github site for the library.

1. *Inserting surfaces and cells.*

```
1    import CardSharp as cs
2    import CardSharpMats as csmat
3    cd = cs.CardDeck()
4    cd.setParticlesList(['p', 'e']) # used for Mode card and IMP string
5    cd.insertMaterialStrings(['Aluminum', 'Air'])
6    # -------------insert surfaces/macros/cells----------------
7    # Inserting surface/cell in one call using convenience functions
8    snSph, cnSph = cd.insertMacroAndCellSphere(name='ASphere', radius=2, shift=(0,0,5))
9    # two surfaceNums in a list and one cellNum is returned
10   snList, cnSphShell = cd.insertMacroAndCellSphereShell(name='ASphericalShell',
11                                           radiusOuter=2, radiusInner=1.5)
12   # Inserting surfaces first, then making a cell by using the surfaceNums
13   snCyl = cd.insertSurface_CylinderAligned(name='ACyl', axis='X', xyz=(0,1,1), radius=1.0)
14   snRpp = cd.insertMacroRpp(name='Rpp', xMinMax=(-5,5), yMinMax=(-6,6), zMinMax=(-7,7))
15   # argument surfaceList does intersection of surfaces
16   cn1 = cd.insertCellString(name='Cylinder', surfaceList=[-snCyl, -snRpp.facets['Xmax'],
17                           -snRpp.facets['Xmin']], shift=(0,0,-5), matName='Aluminum')
18   # manualSurfacesString is alternative to surfaceList, unions, intersection etc
19   manualSurfacesString = '(%s:%s) %s %s'%(-snCyl, -snList[0],
20                                       -snRpp.facets['Xmax'], -snRpp.facets['Xmin'])
21   cn2 = cd.insertCellString(name='jumble', manualSurfacesString=manualSurfacesString,
22                       matName='Aluminum', shift=(0,0,-10))
23   #--------------insert universe-----------
24   # Universe macro number and world/graveyard cell numbers are returned
25   # defines world using automatic cell complement
26   worldSurfaceNum, cellList = cd.insertWorldMacroAndCell(pos=(0,0,0), radius=25,
27                                       worldMat='Air')
28   deckStr = cd.assembleDeck(titleCard='Title card: Test 1')
29   # Save file -------------------------------------------------
30   modelFolder = '../CardSharpOutput/Tests/Temp/'
31   modelFilename = "test.i"
32   cd.saveDeck(modelFolder, modelFilename, deckStr)
```

Lines 1,2 import the library. Line 3 creates a CardDeck object which will hold all the cards needed for the simulation. Line 4 sets the particles for the default importance string for all the cells and the MCNP execution mode. Line 5 inserts materials from the built-in materials dictionary into the card deck. Line 8 inserts a sphere macro and a cell based on that macro with the default position/size. The auto assigned surfaceNum and cellNum are returned. This is a convenience function that combines the insertion of the surface and the cell. Convenience functions do in one call something that would otherwise take two or three calls, but you don't have to use them. Line 10 is another convenience function to insert a spherical shell. It returns the two auto assigned surface numbers (inner/outer) in a list and also the auto assigned cell number. The returned surfaces numbers can be used in more cells using the function insertCellString later in the deck. Line 13 inserts a cylinder macro into the card deck. Line 14 inserts a RPP macro. The auto assigned surface numbers are returned and saved in variable snCyl, snRpp. Line 16 inserts a cell using snCyl and facets of snRpp using only intersections. Line 19 makes a manual surfaces string with unions and intersections. Line 21 inserts a cell using the manual surfaces string. (One can argue that manual generation of the union/intersection string is not much better than the raw MCNP syntax, but since the user still has to learn to debug the MCNP deck, there is no advantage in introducing new syntax. The benefit comes from not having to manage raw cell/surface numbers, being able to choose the order in which surfaces and cells are generated for readability and use Python loops, illustrated later). Line 26, insertWorldMacroAndCell is another convenience function. It inserts a spherical surface to bound the problem region and helps to fill all the undefined space inside using one of two ways. The user can specify a list of intersection surfaces as usual or rely on the convenience function to fill all the undefined space using the cell complement method. (Internally, this is implemented by the library keeping track of all the cells that are inserted in universe=0.) We recommend starting with the cell complement method. If there is reason to believe that the cell complement method results in slow model execution, then one can fall back upon traditional methods to explicitly define the remaining volume regions. Line 28 assembles the whole deck with a title card and returns the whole deck as a multi-line string. Lines 30-32 saves the deck to the input file.

The function to insert a cell into the deck has the following signature:

```
1    insertCellString(name, surfaceList=None, cellComplementList=None,
2                     manualSurfacesString=None,
3                     matName='Void', density=0,
4                     shift=(0,0,0), rotMatrix=None,
5                     impString='', cellNum=None, uni=0, fill=0,
6                     latticeType=None, latticeIndices=None)
7    insertCellLike(name, oldCellNum, newCellNum=None, shift=None,
8                     rotMatrix=None, impString='', uni=0)
```

As can be seen, most arguments have a default values. The user needs to provide only the ones that are necessary. The argument name is for documentation. The surfaceList gives a list of surfaces whose intersection will define the cell's volume. If density is left at zero, the default density from the materials dictionary will be used. If the user provides a shift or rotMatrix, a TRCL string with minimal number of arguments will be appended to the cell. If cellNum=None is passed in, a cell number is auto assigned and a variable holding the cellNum returned. If an integer or string greater like 1 or '1' is passed in as cellNum, it is used as the cellNumber but mixing manually assigned and auto generated cellNumbers can cause errors (unless you establish clear ranges for manual and automatic assignment). Existing cells can be cloned using the LIKE n BUT mechanism using the function shown on Line 7.

*2. Finding/aliasing/inserting materials, adding materials not in library*

```
1    cd = cs.CardDeck()
2    #============MATERIALS START=============================
3    #csmat.matSearch('coba'); return
4    customMatString = """\
5    c      Cobalt 60
6    c      rho = 8.9 g/cc
7    m{}   60000   -1.0   $ Co-60"""
8
9    # insert custom material into database
10   csmat.matInsert(key='Co60', matString=customMatString, defaultDensity=-8.9)
11
12   # Some materials from compendium need a shorter alias
13   csmat.matAddAlias(SodiumIodide02wtThaliumDoped, 'NaI')
14   # insert material into card deck
15   cd.insertMaterialStrings(['Co60', 'Aluminum', 'NaI', 'Air', 'PC', 'FR4PCB'])
16   #============MATERIALS END=============================
```

Line 2 shows how to search the included library for a material with a partial string. If the material is not found in the library, Lines 4-10 show how to add a custom material to the material to the library. Lines 4-7 create a MCNP style string with a placeholder for the material number which will be auto assigned. Line 10 inserts the material into the materials dictionary along with the default density. The key provided is used to refer to the material when inserting a cell. If the cell does not specify a density, the default density from the library is used. Line 13 creates an alias (NaI) to a material in the library with a long or unwieldy name. Line 15 adds materials/aliases from the materials dictionary to the card deck.

*3. TN card for surfaces, loop/nested loop*

```
1    # Creating a hexagonal prism using transforms and a loop
2    intercept = hexagonSideLen * 0.866
3    snList = [] # surface numbers
4    angList = [0, 60, 120, 180, 240, 300]
5    for i,ang in enumerate(angList):
6      rotMatrix=cd.getRotationMatrix(rotationAxis='Z', angleDeg=ang)
7      trNum = cd.insertTRString(name='Hex%d'%(i+1), rotMatrix=rotMatrix)
8      sn = cd.insertSurface_PlaneAligned('HexPlane%d'%(i+1), axis='Y',
```

```
 9                                          D=intercept, trNum=trNum)
10         snList.append(sn)
```

The code above illustrates creating the sides of a hexagonal prism using transforms (TR cards) and a python loop. Line 3 snList is where we collect the surface numbers. Lines 6,7 create a rotation matrix and insert a TR card into the card deck. The auto assigned trNum is returned. Line 8 inserts a plane with the transform into the card deck.

### 4. Universe, Fill, Lattice

```
 1     # Define lattice cell in uni=1
 2     msph, cn1 = cd.insertMacroAndCellSphere(name='sphere', pos=(0,0,0), radius=3,
 3                              matName='Lead', uni=1)
 4     mrpp = cd.insertMacroRpp(name='rpp', xMinMax=(-5,5), yMinMax=(-5,5), zMinMax=(-5,5))
 5     cn2 = cd.insertCellString(name='rpp_cell', surfaceList=[-mrpp, msph],
 6                              matName='Air', uni=1)
 7     # Done with defining cell in uni=1
 8     #---------------------------
 9     # A lattice must be the only thing in its universe,
10     # So we will use an in between universe, uni=2
11     cn = cd.insertCellString(name='rpp_lattice', surfaceList=[-mrpp],
12                              matName='Void', density=0, # this material is ignored
13                              impString='', uni=2, latticeType=1,
14                              latticeIndices=(-1,1,-1,1,-1,1),
15                              fill=[1]*27) # fill with uni=1
16     #---------------------------
17     # Now define a RPP surface to bring the lattice universe into uni=0
18     #(the real world where particles run)
19     mrpp2 = cd.insertMacroRpp(name='rpp', xMinMax=(-5*3,5*3),
20                                          yMinMax=(-5*3,5*3),
21                                          zMinMax=(-5*3,5*3))
22     cn = cd.insertCellString(name='rpp', surfaceList=[-mrpp2],
23                                          fill=2) # fill with uni=2
```

The MCNP method of generating lattices is not the easiest to follow. The library tries to ease the burden, but the user still needs to read the manual and learn the concepts behind the MCNP lattices. The code above illustrates how a lattice cell is defined and a rectangular lattice filled with it.

Lines 1-7 define a rectangular parallelepiped (RPP) region containing a lead sphere, surrounded by air. These are placed into universe 1 using uni=1. Next a lattice is generated in universe 2 using the cell defined earlier. This is done specifying that we are putting the lattice in uni=2, by filling with fill=1 (fill with universe 1), using a latticeType 1 (rectangular) and lattice index ranges for all three axes. Here all three axes have lattice indices going from –1 to +1, i.e. three indices for each axis, giving us 27 lattice cells. And each of the lattice cells is filled with universe 1.

Finally in lines 18-20, the lattice is brought into the real world, which is always uni=0, as dictated by MCNP.

### 5. Source types (or SDEF card)

```
 1     def insertSource_PointWithAngularAndEnergyDistrib(pos=[0,0,0],
```

```
2                                    dirDistrib=None, # None, Bias, Restrict
3                                    vec=[0,1,0], coneHalfAngleDeg=25.8,
4                                    ergDistrib='Discrete', # 'Histogram', 'Continuous'
5                                    eList=[.3, .5, 1.0, 2.5], relFq=[.2, .1, .3, .4],
6                                    par='P', trNum=None)
```

Like other functions in the library, the source functions have many arguments, but most have good default values, and the user only needs to provide the ones relevant to their case. For ex. the above method for inserting the point source does not need any arguments to insert an isotropic point source at the origin, emitting photons with the default energy lines with the default relative fqs. If an anisotropic source or a biased source is needed, set the dirDistrib to 'Restrict' or 'Bias' and set the vec to the correct direction and the coneHalfAngleDeg to the needed angular width. (A biased source still models an isotropic source, but converges faster). Changing the energy distribution is done using the ergDistrib, eList, relFq arguments. The argument par is used to change the emitted particle type and trNum is used to apply a previously inserted geometric transformation number. The trNum can be used to rotate or shift the source to the correct location/orientation.

```
1     def insertSource_BoxWithAngularAndEnergyDistrib(
2                                    xRange=[0,1], yRange=[0,1], zRange=[0,1],
3                                    dirDistrib=None, # None, Bias, Restrict
4                                    vec=[0,1,0], coneHalfAngleDeg=1,
5                                    ergDistrib='Discrete', # 'Histogram', 'Continuous'
6                                    eList=[.3, .5, 1.0, 2.5], relFq=[0, .1, .3, .4],
7                                    rejCell=None, eff=0.01,
8                                    par='P', trNum=None):
```

For the above method to insert a volumetric box source, most of the arguments are the same as the point source described previously. xRange/yRange/zRange define the box shaped region to sample. If sampling an uneven shaped cell enclosed by the box, use the argument rejCell (rejection cell). The argument eff indicates the minimum acceptable sampling efficiency. There are similar methods for a volumetric spherical and volumetric cylindrical source.

### 6. Tallies

```
1     insertF5Tally(tallyNum, pos=(100,0,0), r=0, eList=None, mList=None, par='p')
```

The above shows the signature for inserting a point tally. The tallyNum is not auto assigned as of now. Since MCNP embeds the tally type in the tally number, if a user uses the above method with tallyNum=2, the actual MCNP tally number inserted would be 25. The argument pos gives the coordinates of where the tally would take place. The argument r is the radius of the exclusion zone required for point tallies. The eList gives the energy bins that the tally will contain and mList gives the optional multiplier that will be applied to each bin.

```
1    insertF1Tally(tallyNum, surfInfo, eList=None, mList=None, par='p')
2    insertF2Tally(tallyNum, surfInfo, eList=None, mList=None, par='p')
3    insertF4Tally(tallyNum, cellInfo, eList=None, mList=None, par='p')
4    insertF6Tally(tallyNum, cellInfo, eList=None, mList=None, par='p')
5    insertF7Tally(tallyNum, cellInfo, eList=None, mList=None)
6    insertF8Tally(tallyNum, cellInfo, eList=None, mList=None, par='p,e')
```

The above shows the signatures of functions for inserting surface/cell tally types. Most of the arguments are similar to the ones for the point tally type, except for the surfInfo. This argument can be used in multiple ways. For simple cases, the surfInfo can be a single integer value identifying a surface, or a float with one decimal digit if the surface is one of the multiple surfaces created by a macro. But MCNP also allows a complex scheme of multiple surfaces (or cells) from a hierarchy of universes that can be combined into a single tally. If that is your case, then surfInfo should be passed in as a string representing the surfaces numbers following the rules defined in the MCNP manual.

```
1    insertFIR5Tally(tallyNum, pos=(100,0,0), normVec=(1,0,0),
2                    sMin=-0.02, sMax=0.02, sbins=40, tMin=-0.02, tMax=0.02, tbins=60,
3                    eList=None, mList=None, par='p')
```

That brings us to the FIR5 tally which is different from all the other tally types. This is a radiography tally which sets up a two-dimensional grid on which tallies are generated. The grid and its location/orientation are specified using pos and normVec. The number of bins in the tally are specified using sMin/sMax/sbins and tMin/tMax/tbins arguments (This is MCNP nomenclature).

### 7. Miscellaneous support functions

```
1    insertPhysicsCard(nocoh=0, ides=0, nodop=0, cutn=0.0, cutp=0.001, cute=0.001)
2    insertOutputControlCards(nps=1000, debugN=None, notrn=False)
3    insertIntoCellSection(s)
4    insertIntoSurfaceSection(s)
5    insertIntoSrcSection(s)
6    insertIntoTallySection(s)
7    insertIntoPhysicsSection(s)
8    insertIntoOutputSection(s)
9    assembleDeck(titleCard, surfaceString='Auto', cellString='Auto', trString='Auto',
         matString='Auto', srcString='Auto', tallyString='Auto', physicsString='Auto',
         outputControlString='Auto')
10   tallyNumWType = cd.insertF8Tally(tallyNum=1, cellInfo=cnDet, eList=eList, par='p')
11   specTreatStr = 'FT%d GEB 0.0 0.031196 4.6141'%(tallyNumWType)
12   cd.insertIntoTallySection(specTreatStr)
```

Most of the functions above and their arguments are self-explanatory. The functions insertIntoXXXsection are for inserting cards that are not implemented in the library. Lines 10-12 illustrate this with the GEB special treatment card. The function assembleDeck does not need to be provided anything other than the titleCard. All the other parameters will be taken from the data that the CardDeck collects as the user calls various function/methods. But should the user wish to completely overwrite any section with a custom multi line string, it can be done by providing the appropriate argument.

**IV. Figures showing supported primitives and interesting geometries generated using the library.**

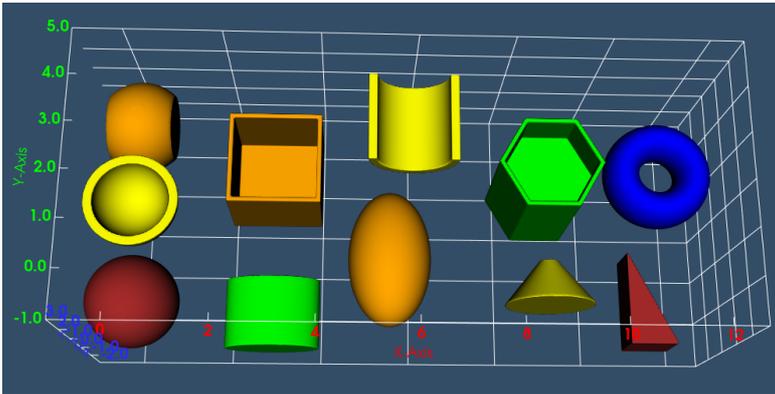

Figure 1. Gxsview-rendered cells created from intersections of a variety of primitive surface types.

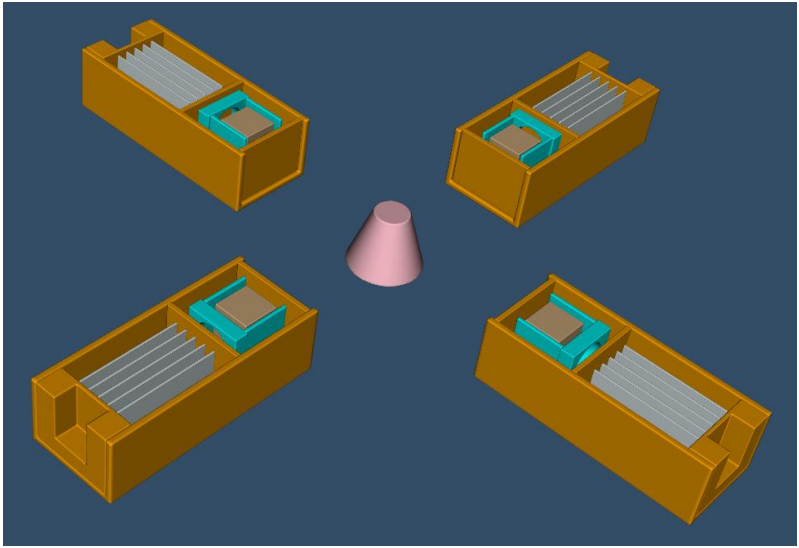

Figure 2. CZT detectors generated using the library. Code on github.

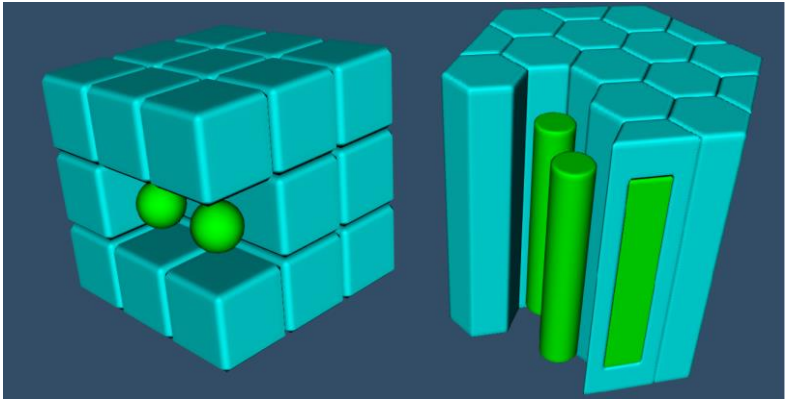

Figure 3: Generated using cubic and hexagonal lattice cards.

**V. CONCLUSION**

We have developed a Python-based library which can greatly simplify and speed up the generation of MCNP input decks. We have tested the library (currently ~3000 lines of code) in multiple projects and found it to be quite useful. More complex sources, tallies and cards will be added in the future. (They can be inserted currently with raw string insertion features provided). Another area for future work is to create/integrate a more comprehensive tally reading module. While the need for periodically referring to MCNP documentation cannot be completely avoided, we believe that the library strikes a good balance by not inventing too much new syntax yet making MCNP input less error prone. It can be useful for both new and experienced MCNP users who have some familiarity with Python. The library code and example codes will be made freely available on github at: https://github.com/pnnl/CardSharpForMCNP. We thank Pacific Northwest National Laboratory for providing access to MCNP to support this work.